# Knot-isomers of Möbius Cyclacene: How Does the Number of Knots Influence the Structure and First Hyperpolarizability?

Hong-Liang Xu,[1] Zhi-Ru Li,[2] Zhong-Min Su,[1,2,*] Feng Long Gu,[3,*] and Kikuo Harigaya[4]

[1] *Institute of Functional Material Chemistry, Faculty of Chemistry, Northeast Normal University, Changchun 130024, Jilin, People's Republic of China,* zmsu@nenu.edu.cn

[2] *State Key Laboratory of Theoretical and Computational Chemistry, Institute of Theoretical Chemistry Jilin University, Changchun, 130023, China,*

[3] *Department of Molecular and Material Sciences, Faculty of Engineering Sciences, Kyushu University, 6-1 Kasuga-Park, Fukuoka, 816-8580, Japan; E-mail:* gu@cube.kyushu-u.ac.jp

[4] *Nanotechnology Research Institute, AIST, Umezono 1-1-1, Tsukuba, Ibaraki 305-8568, Japan; E-mail:* k.harigaya@aist.go.jp

**Abstract**

Four large ring molecules composed by 15 nitrogen-substituted benzene rings, named as "knot-isomers of Möbius cyclacene", i.e. non-Möbius cyclacenes without a knot (**0**), Möbius cyclacenes with a knot (**1**), non-Möbius cyclacenes with two knots (**2**), and Möbius cyclacenes with three knots (**3**), are systematically studied for their structures and nonlinear optical properties. The first hyperpolarizability ($\beta_0$) values of these four knot-isomers structures are 4693 (**0**) < 10484 (**2**) < 25419 (**3**) < 60846 au (**1**). The $\beta_0$ values (60846 for **1**, 10484 for **2** and 25419 au for **3**) of the knot-isomers with knot(s) are larger than that (4693 au for **0**) of the knot-isomer without a knot. It shows that the $\beta_0$ value can be dramatically increases (13 times) by introducing the knot(s) to the cyclacenes structures. It is found that introducing knots to cyclacenes is a new means to enhance the first hyperpolarizability.

Two noticeable relationships between the number of knots and the first hyperpolarizability have been observed. i). the $β_0$ values (60846 for **1** and 25419 au for **3**) of one surface Möbius cyclacene (**1** and **3**) with odd number of knots are larger than that (4693 for **0** and 10484 for **2**) of two surfaces non-Möbius cyclacenes (**0** and **2**) with even number of knots. ii). For the one surface Möbius cyclacenes, the $β_0$ value (60846) for **1** with one knot is larger than that (25419 au) for **3** with three knots.

On the other hand, the largest component of $β_0$ for the four knot-isomer of Möbius cyclacene is alternated among x, y and z for different number of knots. The largest component is $β_z$ for the **0**, after twisting the **0** with the first knot and the second knots, the largest component turns to $β_y$ for the **1** and **2**. The largest component turns back to the $β_z$ for the **3**.

**Introduction**

Since the famous one surface Möbius strip[1] was discovered by German mathematician Möbius in 1858, the special structural[2] and curious chemical and physical[3] characteristics of one surface Möbius strip with a knot have drawn extensive attention of the scientists. For example, for the aromaticity, the Hückel rules for aromaticity (4n+2 electrons) are no longer valid for Möbius annulenes, but the Möbius ring with 4n π electrons is aromatic. For the magnetism, the unusual ring currents in Möbius annulenes are also particularly interesting.

Nonlinear optics[4] (NLO) develops very quickly in the past two decades. Much effort has been devoted to find the important influencing factors which can lead to a significant increase in the first hyperpolarizability and to design new type NLO materials. Theoretical investigations[5] play an important role for the new high-performance NLO materials' discovery. But the NLO properties for the Möbius systems are seldom studied.

The framework shape effect on the first hyperpolarizability, however, has been investigated by comparing non-Möbius cyclacenes and Möbius cyclacenes.[6] It was shown that twisting a knot of non-Möbius (normal) cyclacenes to form Möbius cyclacenes the first hyperpolarizability is decreased from 1049 to 393 au. Whether the first hyperpolarizability can be increased when adding a knot into non-Möbius cyclacenes to form Möbius cyclacenes? This work is trying to answer this question.

**Computational Details**

The optimized geometric structures of four large knot-isomers of Möbius cyclacene with all real frequencies are obtained by using the density functional theory (DFT) B3LYP/6-31G(d) level.

Champagne and Nakano pointed out that, for a medium-size system, p-quinodimethane, the BHandHLYP method can also reproduce the (hyper)polarizability values from the more sophisticated the single, double, and perturbative triple excitation coupled-cluster [CCSD(T)].[7] For the Möbius cyclacenes with seven nitrogen-substituted benzene rings and relative systems,[6] the satisfying results of BhandHlyp $\beta_0$ value are obtained (see Supporting Information). Thus, the first (hyper)polarizabilities are evaluated for the four large knot-isomers of Möbius cyclacene in the present work at BhandHlyp/6-31+G(d) level. The magnitude of the applied electric field is chosen as 0.001 au for the calculation of the (hyper)polarizabilities.

The polarizability ($\alpha_0$) is defined as follows:

$$\alpha_0 = \frac{1}{3}(\alpha_{xx} + \alpha_{yy} + \alpha_{zz}) \tag{1}$$

The static first hyperpolarizability is noted as:

$$\beta_0 = (\beta_x^2 + \beta_y^2 + \beta_z^2)^{1/2} \tag{2}$$

where $\beta_i = \frac{3}{5}(\beta_{iii} + \beta_{ijj} + \beta_{ikk}), i, j, k = x, y, z$.

All of the calculations were performed with the GAUSSIAN 03 program package.[8] The dimensional plots of molecular orbitals were generated with the GaussView program.[9]

**Results and Discussions**

*A. Equilibrium Geometries*

The optimized geometric structures of four knot-isomers of Möbius cyclacene with all real frequencies are shown in Figure 1.

The four knot-isomers of Möbius cyclacene are named by the knot number (**0, 1, 2** and **3**), and the dihedral angles $C_n$-C-C-C (n=1, 2, 3… 15) are denoted in Figure 1a. In Figure 1, the fifteen nitrogen substituted [15]cyclacene is two surface non-Möbius cyclacene **0** without a knot. Twisting the first knot, the one surface Möbius cyclacene **1** with one knot is formed. Further twisting second knot, we obtained the two surfaces non-Möbius cyclacene **2** with two knots. Further twisting third knot, the new one surface Möbius cyclacene **3** with three knots is obtained.

How does the number of knots influence the structure? It is found that the number of the dihedral angle $C_n$-C-C-C peaks consists with the knot number (see Figure 2). From Table I and Figure 2, the **0** has the same the dihedral angles $C_n$-C-C-C (0.005~0.035°). Twisting first knot to form one surface Möbius cyclacenes (**1**), the dihedral angle peak is formed on the $C_{13}$-C-C-C dihedral angle (31.693°). Twisting second knot form two surfaces cyclacenes (**2**), two peaks are formed on the $C_5$-C-C-C (31.061°) and $C_{12}$-C-C-C dihedral angle (22.635°). Twisting third knot to form another Möbius cyclacenes (**3**), three peaks are formed on the $C_5$-C-C-C (29.962°), $C_9$-C-C-C (32.207°) and $C_{14}$-C-C-C dihedral angle (38.134°).

*B. The Static First Hyperpolarizabilities*

The electric properties of **0, 1, 2** and **3** calculated at the BHandHLYP/6-31+G(d) level are given in Table II. From Table II, the order of polarizability ($α_0$) is 927.17 (**2**) < 960.48 (**0**) < 1077.67 (**3**) <

1088.85 (**1**) au. The knot number effect on the polarizability for knot-isomers of Möbius cyclacene is shown: the $\alpha_0$ values (1088.85 for **1** and 1077.67 au for **3**) of one surface Möbius cyclacene (**1** and **3**) with odd number of knots are larger than that (960.48 for **0** and 927.17 au for **2**) of two surfaces non-Möbius cyclacenes (**0** and **2**) with even number of knots.

Especially, the relationships between the first hyperpolarizability and knot number are investigated. The order of $\beta_0$ values is 4693 (**0**) < 10484 (**2**) < 25419 (**3**) < 60846 au (**1**). Comparing these $\beta_0$ values, we find that the $\beta_0$ values (60846 for **1**, 10484 for **2** and 25419 au for **3**) of the structures with knot(s) are larger than that (4693 au for **0**) of the structure without knot (see Figure 3). It shows that the $\beta_0$ value can be increased by introducing the knot(s) to the cyclacene, which is new factor to enhance the first hyperpolarizability.

Two noticeable relationships between the knot number and the first hyperpolarizability have been observed.

i). The $\beta_0$ values (60846 and 25419 au) of one surface Möbius cyclacenes (**1** and **3**) with odd number of knots are large that (4693 and 10484 au) of two surfaces non-Möbius cyclacenes (**0** and **2**) with even number of knots.

ii). For the one surface Möbius cyclacenes, the $\beta_0$ value (60846) for **1** with one knot is larger than that (25419 au) for **3** with three knots, which shows that the structure with small knot number of **1** has large $\beta_0$ value.

Among the four knot-isomers of Möbius cyclacenes, the one surface Möbius cyclacene with a knot has the largest $\beta_0$ values (60846 au), which can compare with that of other high NLO systems. For example, the known electrides $(HCN)_nLi$ [10a], Li@calix[4]pyrrole [10b] (the range of the $\beta_0$ values is 3385 ~ 15682 a.u.), and the large donor-acceptor polyenes systems[11a] (the range of the $\beta_0$ values is 8818 ~

152502 a.u.) as well the organometallic system *cis*-[RuII(NH$_3$)$_4$(2-PymQ$^+$)$_2$][PF$_6$]$_4$ (34487 a.u).[11b]

The first hyperpolarizability can be estimated by the two-state approximation,[12]

$$\beta_0 \propto \Delta\mu_{ge} \frac{\mu_{ge}^2}{E_{ge}^2}, \quad \mu_{ge} = \mu_{ee} - \mu_{gg} \quad (3)$$

where the subscript "g" indicates the ground state and the subscript "e" indicates the charge-transfer excited state. $\Delta\mu_{ge}$ is the dipole moment difference, $\mu_{ge}$ is the transition dipole moment, and $E_{ge}$ is the transition energy. From the two-level expression (eq. 3), it is obvious that the transition energy is the decisive factor in the first hyperpolarizability. A simple approximation is to represent the ground and excited states using the HOMO (H) and LUMO (L) orbital energies,[13] respectively, so the $E_{ge}$ is denoted $\varepsilon_{gap}$ in the present work.

From Eq 3, the $\beta_0$ is inversely proportional to the second power of the transition energy $\varepsilon_{gap}$. The order of the $\varepsilon_{gap}$ is 2.137 (**0**) > 1.459 (**1**) < 2.092 (**2**) > 2.014 (**3**) eV, which explains the relationship between the knot number and $\beta_0$. Among the four knot-isomers of Möbius cyclacene, the $\varepsilon_{gap}$ value (1.459 for **1**, 2.092 for **2** and 2.014 eV for **3**) of structures with knot(s) are smaller than that (2.137 eV for **0**) of structure without knot. And the $\varepsilon_{gap}$ of one surface Möbius cyclacene (**1** and **3**) with odd number of knots are smaller than that of two surfaces cyclacene (**0** and **2**) with even number of knots. For the one surface Möbius cyclacenes, the $\varepsilon_{gap}$ (1.459) for **1** (one knot) smaller than that (2.014 eV) for **3** (three knots). Interestingly, Figure 3 shows the shape of the **1/$\varepsilon^2_{gap}$** curve is very similar to that of $\beta_0$ curve.

On the other hand, it is observed that the largest component of $\beta_0$ for the four knot-isomer of Möbius cyclacene is alternated among x, y and z for different number of knots. The largest component is $\beta_z$ for the **0**, after twisting the **0** with the first knot and the second knots, the largest component turns to $\beta_y$ for the **1** and **2**. The largest component turns back to the $\beta_z$ for the **3**.

From Eq 3, the largest component of $\Delta\mu_{ge}$ is related to the largest component of $\beta_0$. From Figure 4, the HOMO of **0** is mainly located on the un-substituted side, and LUMO of **0** is mainly located on the N-substituted side. Thus, for the HOMO-LUMO transition of **0**, the charge transfer and $\Delta\mu_{ge}$ along with the z-axis from un-substituted side to the N-substituted side. So the largest component of $\beta_0$ is $\beta_z$. While For **1,** the HOMO is mainly located on the side of positive y-axis, and LUMO is mainly located on the side of negative y-axis. The transition of **1**, the charge transfer and $\Delta\mu_{ge}$ from side of positive y-axis to the side of negative y-axis, and thus the largest component of $\beta_0$ is $\beta_y$. Similar to **1**, the largest component of $\beta_0$ for **2** is also $\beta_y$. Analogically to **0**, the largest component of $\beta_0$ for **3** is $\beta_z$.

**Conclusions**

In the present work, we have obtained a valuable description of the knot effect on the static (hyper)polarizabilities for the four big knot-isomers of Möbius cyclacene, the $\beta_0$ values (60846 for **1**, 10484 for **2** and 25419 au for **3**) of isomer cyclacenes with knot(s) are larger than that (4693 au for **0**) of isomer cyclacenes without a knot. It shows that the $\beta_0$ value can be increased by twisting the knot(s), which is a new factor to enhance the first hyperpolarizability.

Two noticeable relationships between the knot number and the first hyperpolarizability have been observed. i). The $\beta_0$ values (60846 for **1** and 25419 au for **3**) of one surface Möbius cyclacene (**1** and **3**) with odd knot are large that (4693 for **0** and 10484 for **2**) of two surfaces non-Möbius cyclacenes (**0** and **2**) with even knot. ii). For the one surface Möbius cyclacenes, the $\beta_0$ value (60846) for **1** with one knot larger than that (25419 au) for **3** with three knots.

On the other hand, the largest component of $\beta_0$ for the four knot-isomer of Möbius cyclacene can be turning with changing of knot number: the largest component is $\beta_z$ for the **0**, after twisting the **0** with the first knot and the second knots, the largest component turns to $\beta_y$ for the **1** and **2**. The largest component turns back to the $β_z$ for the **3**. As a result, our investigation may offer new way to design new NLO compounds.

**Acknowledgment**

This work was supported by the National Natural Science Foundation of China (No. 20573043, 20773046 and 20703008), and Chang Jiang Scholars Program (2006), Program for Changjiang Scholars and Innovative Research Team in University (IRT0714).

**Supporting Information Available:**

Complete ref 8 and optimized Cartesian coordinates for four big knot-isomers of Möbius cyclacene. This material is available free of charge via the Internet at http://pubs.acs.org.

**Table I.** The dihedral angles $C_n$-C-C-C (n=1, 2, 3… 15) for Knot-isomers of Möbius Cyclacene **0, 1, 2** and **3**.

| $C_n$-C-C-C[a] | 0 | 1 | 2 | 3 |
|---|---|---|---|---|
| 1 | 0.008 | 8.408 | 8.237 | 28.206 |
| 2 | 0.006 | 2.458 | 10.312 | 27.513 |
| 3 | 0.010 | 0.061 | 18.487 | 28.323 |
| 4 | 0.007 | 0.676 | 27.37 | 28.47 |
| 5 | 0.005 | 0.455 | 31.061 | 29.962 |
| 6 | 0.030 | 3.093 | 27.367 | 27.903 |
| 7 | 0.006 | 3.029 | 18.48 | 27.259 |
| 8 | 0.035 | 0.476 | 10.314 | 30.783 |
| 9 | 0.010 | 5.388 | 8.292 | 32.207 |
| 10 | 0.012 | 14.573 | 12.665 | 29.790 |
| 11 | 0.018 | 24.322 | 18.884 | 27.824 |
| 12 | 0.026 | 31.106 | 22.635 | 30.608 |
| 13 | 0.015 | 31.693 | 22.604 | 36.177 |
| 14 | 0.008 | 26.000 | 18.806 | 38.134 |
| 15 | 0.009 | 17.129 | 12.564 | 33.382 |

[a] see Figure 1a.

**Table II.** The values of the $\alpha$, $\beta_0$, LUMO and HOMO at BhandhLYP/6-31+G(d) level for Knot-isomers of Möbius Cyclacene **0, 1, 2** and **3**.

|  | **0** | **1** | **2** | **3** |
|---|---|---|---|---|
| $\alpha$(au) | 960 | 1089 | 927 | 1078 |
| $\beta_x$(au) | -29 | -7885 | -37 | 9886 |
| $\beta_y$(au) | -11 | 59949 | 10484 | 4966 |
| $\beta_z$(au) | 4693 | 6756 | -2 | 22885 |
| $\beta_0$(au) | 4693 | 60845 | 10484 | 25419 |
| HOMO (eV) | -6.081 | -5.709 | -6.086 | -6.313 |
| LUMO (eV) | -3.944 | -4.250 | -3.994 | -4.299 |
| $\varepsilon_{gap}$ (eV) | 2.137 | 1.459 | 2.092 | 2.014 |
| $1/\varepsilon^2_{gap}$(au) | 162.05 | 347.82 | 169.26 | 182.52 |

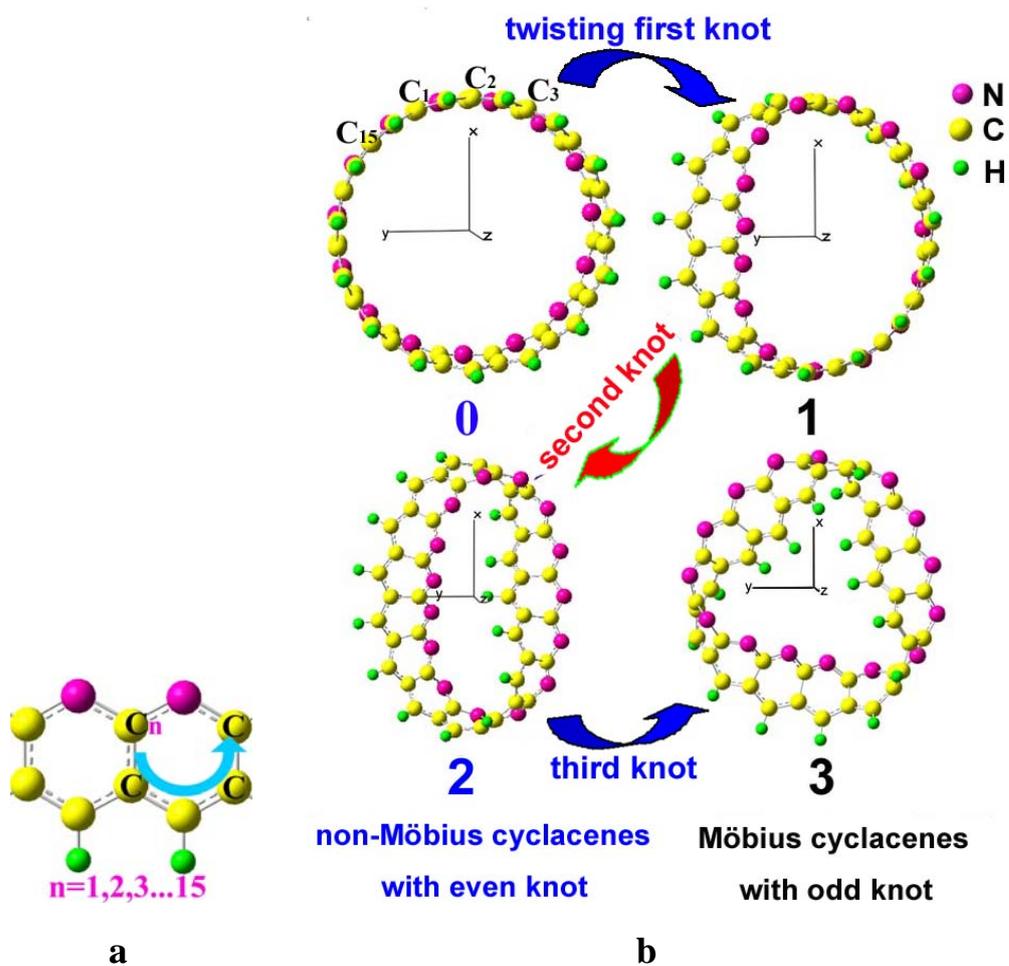

**Figure 1.** The dihedral angles $C_n$-C-C-C (n=1, 2, 3… 15) (azury arrowhead) (**a**). The optimized structures of the the non-Möbius cyclacenes without a knot (**0**), the Möbius cyclacenes with a knot (**1**), new non-Möbius cyclacenes with two knots (**2**) and new Möbius cyclacenes with three knots (**3**) nitrogen-substituted polyacenes (**b**).

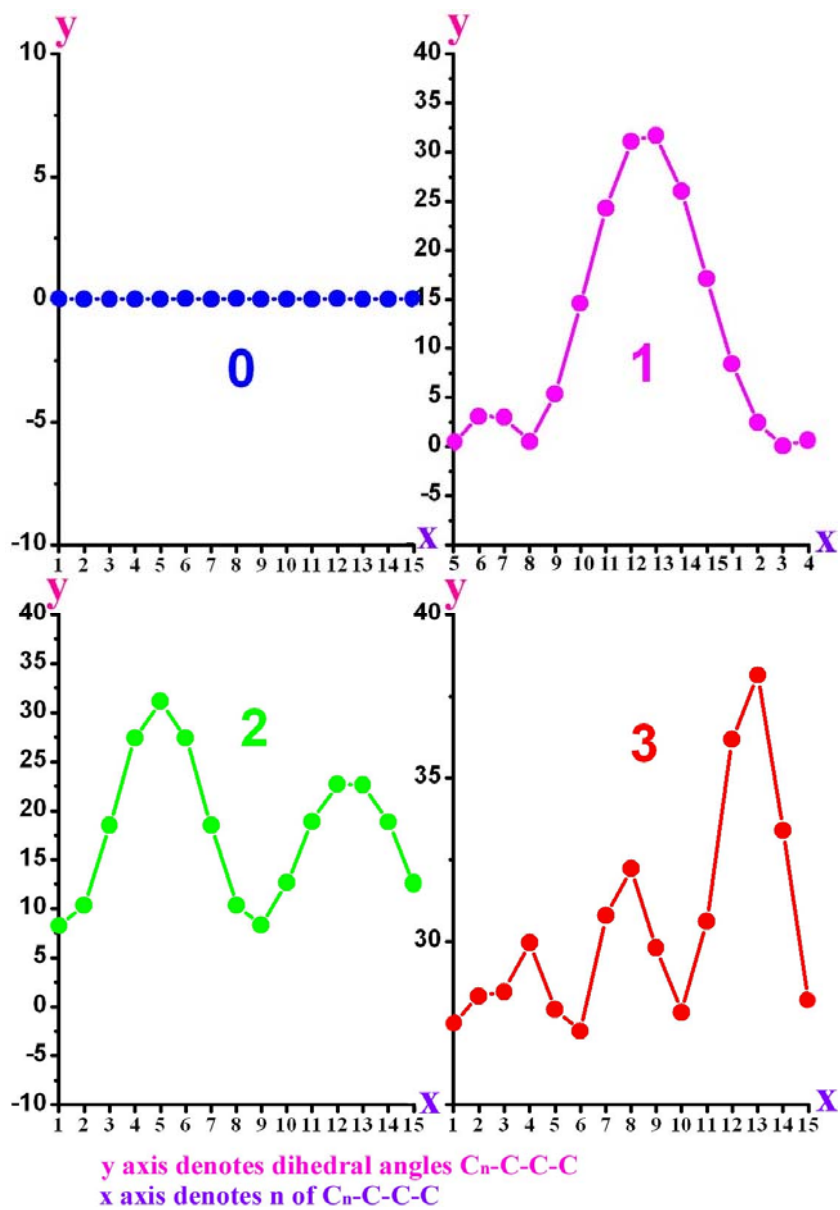

**Figure 2.** The dihedral angles C$_n$-C-C-C (n=1, 2, 3…15) for Knot-isomers of Möbius Cyclacene **0, 1, 2** and **3.**

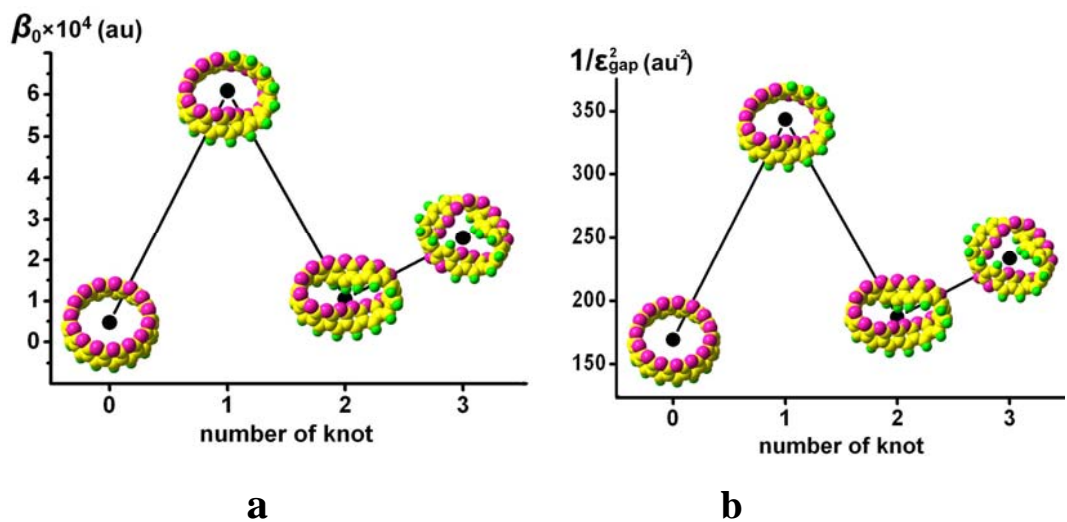

**Figure 3.** The relationship (**a**) between the first hyperpolarizability and knot number, the relationship (**b**) between the reciprocal of $\varepsilon^2_{gap}$ and knot number.

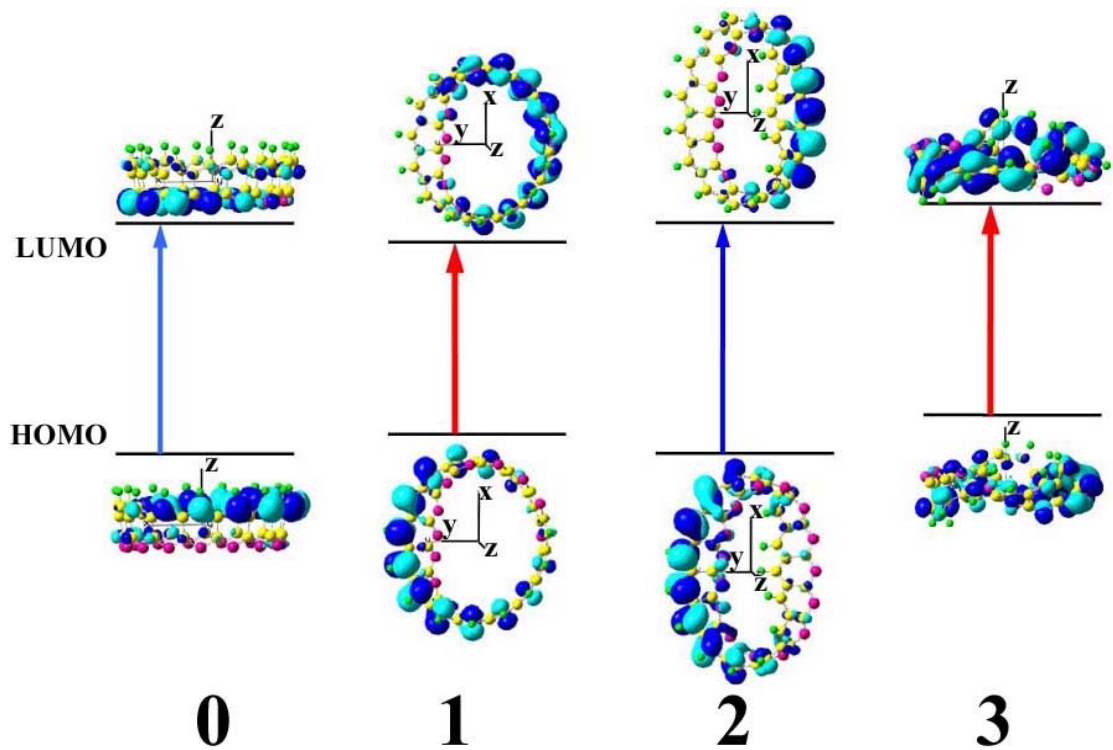

**Figure 4.** The HOMO and LUMO of non-Möbius cyclacenes (blue lines) and Möbius cyclacenes (red lines).

TOC

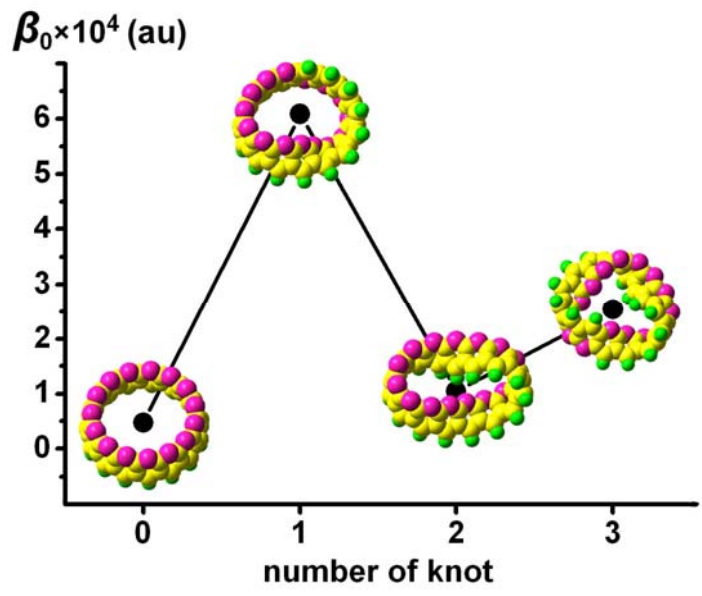